\documentclass[11pt,a4paper,eps,nofootinbib,superscriptaddress]{revtex4}

\newcommand{\be}{\begin{equation}}
 \newcommand{\ee}{\end{equation}}
\newcommand{\bea}{\begin{eqnarray}}
\newcommand{\eea}{\end{eqnarray}}
\newcommand{\f}{\frac}

\newcommand{\s}{s}
\newcommand{\cplus}{\dot+}
\newcommand{\ts}{\left(}
\newcommand{\td}{\right)}
\newcommand{\qs}{\left[}
\newcommand{\qd}{\right]}

\newcommand{\nn}{\nonumber}
\newcommand{\x}{{\bf x }}

 \newcommand{\kM}{{$\kappa$-Minkowski}}

\usepackage{amssymb}

\usepackage{amsmath}

\usepackage{graphicx}

\usepackage{amsfonts}

\begin{document}

\title{ Interplay between curvature and Planck-scale effects\\ in astrophysics and cosmology}

\author{Antonino MARCIAN\`{O}}
\email{antonino.marciano@cpt.univ-mrs.fr}
\affiliation{Dipartimento di Fisica, Universit\`{a} di Roma ``La Sapienza'', and sez. Roma1 INFN,  P.le A. Moro 2, 00185 Roma, EU}
\affiliation{Centre de Physique Th\'eorique de Luminy, case 907, f-13288 Marseille, EU}
\author{Giovanni AMELINO-CAMELIA}
\affiliation{Dipartimento di Fisica, Universit\`{a} di Roma ``La Sapienza'', and sez. Roma1 INFN,  P.le A. Moro 2, 00185 Roma, EU}
\author{Nicola Rossano BRUNO}
\email{r.bruno@bnrenergia.it}
\affiliation{Centro Studi e Formazione sulle Energie Rinnovabili, BNR Energia Srl, Via Costabella 34/36, 00195 Roma, EU}
\author{Giulia GUBITOSI}
\affiliation{Dipartimento di Fisica, Universit\`{a} di Roma ``La Sapienza'', and sez. Roma1 INFN,  P.le A. Moro 2, 00185 Roma, EU}
\author{ Gianluca MANDANICI}
\email{gianluca.mandanici@istruzione.it}
\author{Alessandro MELCHIORRI}
\affiliation{Dipartimento di Fisica, Universit\`{a} di Roma ``La Sapienza'', and sez. Roma1 INFN,  P.le A. Moro 2, 00185 Roma, EU}

\maketitle

\baselineskip 12pt plus .5pt minus .5pt \pagenumbering{arabic} %

\begin{center}
\textbf{ABSTRACT}
\end{center}

\noindent Several recent studies have considered the implications
for astrophysics and cosmology of some possible nonclassical
properties of spacetime at the Planck scale. The new effects,
such as a Planck-scale-modified energy-momentum (dispersion) relation,
are often inferred from the analysis of some quantum versions of Minkowski spacetime,
and therefore the relevant estimates depend heavily on the assumption
that there could not be significant interplay between Planck-scale and curvature effects.
We here scrutinize this assumption, using as guidance
a quantum version of de Sitter spacetime
 with known In\"on\"u-Wigner contraction to a  quantum Minkowski spacetime.
 And we show that, contrary to common (but unsupported) beliefs,
the interplay between Planck-scale and curvature effects can be significant.
Within our illustrative example, in the Minkowski limit the quantum-geometry
deformation parameter is indeed given by the Planck scale,
while in the de Sitter picture the parameter of quantization
of geometry depends both on the Planck scale and the
curvature scalar. For the much-studied case of Planck-scale
effects that intervene in the observation of gamma-ray bursts
we can estimate the implications of ``quantum spacetime curvature''
within robust simplifying assumptions.
For cosmology at the present stage of the development of the relevant
mathematics one cannot go beyond semiheuristic reasoning, and we here
propose a candidate approximate description of a
quantum FRW geometry, obtained by patching together
pieces (with different spacetime curvature) of our quantum de Sitter.
This semiheuristic picture, in spite of its limitations,
provides rather robust evidence that in the early Universe
the interplay between Planck-scale and curvature effects
could have been particularly significant.

\newpage
\pagenumbering{arabic}
 \pagestyle{plain}
%

\section{Introduction}
For many decades~\cite{stachelHISTO,gacQM100nature} progress in the study of the quantum-gravity
problem was obstructed by the extreme mathematical complexity
of the most promising theories
of quantum gravity, resulting in a
debate that was confined at the level of comparison
of mathematical and conceptual features.
At least for one aspect of the quantum-gravity problem, the one that concerns the
possibility that spacetime itself might have to be quantized,
the nature of the debate started to change in the second half
of 1990s when it was  established that some scenarios for the quantization of spacetime
have implications for spacetime symmetries,  which have then been studied
focusing mainly on the aspects of modification of
the classical-spacetime  ``dispersion" relation between energy
and momentum of a microscopic particle.
These developments have also motivated a rather large effort
on the side of
phenomenology (see, {\it e.g.},
Refs.~\cite{grbgac,astroBiller,kifune,ita,aus,gactp,tedOLDgood,emnPLB2009,gacSMOLINprd,fermiNATURE,atomINTprl})
looking for ways to gain experimental insight on this hypothesis
both in laboratory experiments and, even more frequently, using astrophysics
observatories.

Cosmology has so far played only a relatively marginal role in this
 phenomenology research effort,
 but it appears likely that this might change in the not-so-distant future.
This expectation originates from the fact that many cosmological observations
 reflect the properties of the Universe at very early times, when the typical
 energies of particles were significantly closer to the Planck scale than the energies presently
 reached in our most advanced particle accelerators. Moreover,
 the particles studied in cosmology have typically travelled ultra-long (``cosmological") distances,
 and therefore even when they are particles of relatively low energies they could be affected
 by a large accumulation of the effects of the ``space-time quantization", which is one of the
 most common expectations emerging from quantum-gravity research.

We are here mainly concerned with a key assumption that is commonly made
in the few studies of quantum-spacetime effects for cosmology that have
been produced so far
(see, {\it e.g.}, Refs.~\cite{jurekcosmo,chongsun,infla1,infla2,infla3,maguREV,birefGiulia}).
This is basically the assumption that the quantum-spacetime effects could be safely
estimated in quantum versions of Minkowski spacetime, and then inserted ``by hand"
as new features for the analysis in cosmology, which of course is not formulated in
Minkowski spacetime.
For example, for what concerns the energy-momentum relation,
one essentially assumes that, if in the Minkowski limit the
energy-momentum relation is of the
type\footnote{In this work we set $c=\hbar=1$.} $m^2=P^{\mu}  \eta_{\mu \nu} P^{\nu}
+F_{flat}(L_{p},P^{\alpha})$, in cases with metric $g_{\mu \nu}$ ($\neq \eta_{\mu \nu}$)
one could still write
\be
m^2=P^{\mu} g_{\mu \nu}P^{\nu}+F_{flat}(L_{p},P^{\alpha}), \label{nuga}
\ee
with the same deformation function $F_{flat}(L_{p},P^{\alpha})$.

We here investigate this issue of the interplay between curvature and Planck-scale
effects within the framework that was introduced for these purposes in Ref.~\cite{kodadsr}
(also see Ref.~\cite{antoLAUREA}),
which advocated the study of a specific example of
quantum de Sitter (dS) spacetime, with known Inonu-Wigner contraction
to a much-studied quantum Minkowski spacetime.
We find that the interplay between curvature and Planck-scale effects
is very significant, and in particular our analysis produces
candidates for relations of the type
\be
m^2=P^{\mu}g^{\Lambda}_{\mu \nu}P^{\nu}+F(\Lambda,L_{p},P^{\alpha}) , \label{Lillo}
\ee
where $g^{\Lambda}_{\mu \nu}$ is the dS metric for cosmological constant $\Lambda$.
The significance of the interplay between curvature and Planck-scale effects
admits in our framework a particularly straightforward description:
our quantum version of dS spacetime is dual to a Hopf algebra
whose characteristic parameter is dimensionless. So the only opportunities
for the Planck scale to appear in the description of the structure
of our quantum spacetime necessarily involves expressing this dimensionless
parameter in terms of the Planck scale and of the only other dimensionful scale
present in the framework, which is indeed the curvature scalar.

While our main technical findings concern a candidate for a quantum dS spacetime,
we argue that at least at a semi-heuristic/semi-quantitative level they are valuable
also for some (yet to be formalized) quantum versions of FRW geometries.
We structure this aspect of our thesis by introducing
 an approximate description of a quantum FRW geometry,
obtained by patching together pieces (with different curvature) of our
 quantum dS.
The quantum-deformation parameter characteristic
of our setup must be specified as a function of
the Planck scale and of the ``effective dS-patch curvature",
and we find that different formulations of this relation
(all with the same Minkowski limit!) lead to very different
descriptions of the path of massless particles. We therefore provide an explicit
example of the significance of the interplay between curvature and
Planck-scale effects in cosmology.

In preparation for the main parts of the analysis,
in the next section we briefly review some well-known
aspects of the classical dS spacetime, mainly establishing
 notation to be used in the following.
 Then in Section~\ref{q-dS} we introduce our quantum version
 of dS spacetime, a ``q-dS spacetime", and its contraction
 to the $\kappa$-Minkowski noncommutative spacetime. $\kappa$-Minkowski is
 a relevant example since it has inspired some of the studies considering
 Planck-scale effects in astrophysics and cosmology. Our q-dS spacetime is a natural
 generalization of $\kappa$-Minkowski to the case of a constant-curvature
 maximally-symmetric spacetime. We work mostly (as one often does also in
 dealing with $\kappa$-Minkowski) using a dual description of our q-dS spacetime
 that relies on an associated symmetry Hopf algebra.
The relevant mathematics is not yet fully developed for the 3+1D case,
and therefore  we find convenient
 to consider primarily the cases of 2+1 and 1+1 spacetime dimensions.
This is not a key limitation in light of the objectives of our analysis:
rather than aiming for detailed quantitative results,
we are mainly interested in exposing the presence of some interplay between
Planck-scale effects and curvature, illustrating some of the typical structures
to be expected for this interplay.

In Section~\ref{astrop} we mainly argue that our results have
implications that are significant even for cases in which the curvature scalar
is constant, because we find that some observables, such as
the distance travelled
by a massless particle in a given time interval,
depend on the Planck scale in measure that depends strongly on the curvature scalar.
This point is at least semi-quantitatively relevant for certain observations
in astrophysics, particular the ones that concern sources that are not too distant,
close enough for the time variation of the curvature scalar to be negligible
at least at a first level of analysis.
But we expect that the interplay between curvature and Planck-scale
effects should acquire even more significance
in FRW-like geometries, with their associated time dependence of the curvature scalar,
and we set up our case by first
noticing, in Section~\ref{classlice}, that at the classical-spacetime level of analysis,
 one can obtain a good description of some aspects of FRW spacetimes
by viewing these spacetimes
as an ensemble of patches of dS spacetimes. The intuition gained in Section~\ref{classlice}
then provides guidance for the analysis reported in Section~\ref{sec:qFRW},
which is centered on the working
assumption that one could get a description of a ``q-FRW spacetime"
by combining patches of q-dS spacetime.
Section~\ref{out}
 offers a few closing remarks on the outlook of this research area.

\section{Preliminaries on classical dS space-time}\label{sec:dS}
In preparation for our analysis it is useful to review some aspects of the classical dS
spacetime, especially the description of its symmetries, the associated conserved charges,
and a recipe for obtaining the path of a massless particle that relies primarily on
the conserved charges.

Our notation is such that the Einstein equation, with cosmological constant $\Lambda$,
is written as
\be
R_{\mu\nu}-\frac{1}{2}g_{\mu\nu}R+\Lambda g_{\mu\nu}=-8\pi G T_{\mu\nu} \label{Paolina}
\ee
where $R_{\mu\nu}$ is the contraction by the metric $g_{\mu\nu}$ of the Riemann tensor, $R$
is the Ricci scalar, and $T_{\mu\nu}$ is
the energy-momentum tensor.

In comoving coordinates the dS solution in 3+1D takes the form\footnote{For consistency with standard
conventions used in astrophysics, here and in the following two sections
we choose a time normalization such that in dS
spacetime $a_{dS}=1$ at the present time, {\it i.e.}
we set the present time to zero, past time to be negative and future
time to be positive. We warn our readers that for Sections~\ref{classlice}-\ref{sec:qFRW}
we shall turn to a different choice of time normalization, for consistency with
the one preferred in most applications in cosmology.}
\bea
ds^2=dt^2-a_{dS}^{2}(t)(dx^2+dy^2+dz^2),\qquad {\rm with} \qquad a_{dS}(t)=e^{\,H t }. \label{eq:FRW}
\eea
It is a solution of (\ref{Paolina}) in empty space ($T_{\mu\nu}=0$) with $\Lambda = 3 H^2$, but
it can also be obtained in various other ways, for example
as a solution of the Einstein equation without
a cosmological term ($\Lambda =0$) when
the energy-momentum tensor is the one for a perfect fluid with energy
density $\varrho=3 H^2 / (8 \pi G)$ and constant pressure $p=-3 H^2 / (8 \pi G)$.

The dS solution, can be viewed as a particular FRW
(Friedmann Robertson Walker) solution, in which the characteristic
time-dependent conformal factor $a(t)$ of FRW solutions
takes the form $a_{dS}(t)=e^{\, H t }$. As for all FRW solutions
the spatial line element (in comoving coordinates) $ dl^2=a^{2}(t)(dx^2+dy^2+dz^2)$
is such that the distance between two spatial points
grows with time.
The geodesics are orthogonal to the space like surface,
and the time $t$ is the proper time for different observers in the Universe expansion.

The dS solution is a  constant-curvature spacetime, so the Riemann curvature tensor
 is completely determined by the Ricci scalar $R$ through the
 relation $R_{\mu\nu\alpha\beta}=\frac{R}{12}(g_{\mu\alpha}g_{\nu\beta}-g_{\mu\beta}g_{\nu\alpha})$.
 Since the  Ricci scalar
is related to $H$ by $R= 12 H^2$, it is clear that the constant $H$ suffices in order to specify
the curvature of dS spacetime.

dS spacetime is conformally-flat ,{\it i.e.} the metric is obtained by a conformal transformation
from the Minkowski metric, and of course one obtains the Minkowski spacetime in the
limit $H \rightarrow 0$. In the 3+1D case it can be described as a surface in a five-dimensional
spacetime (with signature $\{-1,1,1,1,1\}$) characterized by the requirement
\be
-z^2_{0}+z^2_{1}+z^2_{2}+z^2_{3}+z^2_{4}=H^2 ~. \label{palloncino}
\ee
The $SO(4,1)$ symmetries transformations of the 3+1D dS spacetime
leave unchanged the bilinear form (\ref{palloncino}), and can be viewed from the perspective
of the embedding 5D spacetime as the 10 rotations which leave invariant
the surface (\ref{palloncino}).

Both the symmetry generators $G_{i}$ (with $i=0,1,,..9$)
and the associated charges $\Pi_{i}$ (conserved along a geodesic line)
can be described in terms of the Killing vectors $\xi_{i}^{\mu}$ of the metric:
\bea
G_{i}=\xi_{i}^{\mu}\partial_{\mu},\\
\Pi_{i}=\xi_{i}^{\mu}p_{\mu}, \label{Pierro}
\eea
where the four-vector $p^{\mu}$, the energy-momentum measured by free-falling observers, is given for a test particle of mass $m$ by
\be
p^{\mu}=m \frac{dx^{\mu}}{d\tau} ~
\ee
along a geodesic with
affine parameter $\tau$.

The Killing vectors
of the metric (\ref{eq:FRW}) are given by
\bea
&&\xi_{P_0}=(1,-H\vec{x}) ~,~~~ \xi_{P_1}=(0,1,0,0) ~,~~~ \xi_{P_2}=(0,0,1,0) ~,~~~
\xi_{P_3}=(0,0,0,1) ~, \nn\\
&&\xi_{N_1} = \ts x, \f{1-e^{-2Ht}}{2 H} -\f{ H}{2} (x^2-y^2-z^2  ), -H x y, -H x z \td   ,\nn\\
&&\xi_{N_2} = \ts y,-H x y, \f{1-e^{-2Ht}}{2 H} -\f{ H}{2} (y^2-x^2-z^2  ),  -H y z \td ,\nn\\
&&\xi_{N_3} = \ts z,-Hxz,-Hyz, \f{1-e^{-2Ht}}{2 H} -\f{ H}{2} (z^2-x^2-y^2  ) \td,\nn\\
&&\xi_{J_1} = ( 0,0,z,-y) ~,~~~   \xi_{J_2} = ( 0,-z,0,x) ~,~~~ \xi_{J_3} = ( 0,y,-x,0) ~.   \label{Kv}
\eea
We are labelling the symmetry generators in a way that refers to the Poincar\'e generators to which they
reduce in the $H \rightarrow 0$ limit.
The generators $G_0\equiv P_0$, $G_i\equiv P_i,\, i=1,2,3$ describe
generalized time-like and space-like translations; the $G_{N_i}\equiv N_i$ are dS
boosts,
and finally $G_{J_i}\equiv J_i$ are rotations. The generators of course close
the $SO(4,1)$ dS classical (Lie) algebra
\bea
&&[P_0 , P_i] = H P_i ~, \qquad [P_0, N_i] = P_i - H N_i ~, \qquad [P_0, J_i] = 0 ~,\nn\\
&&[P_i,P_j]=0 ~, \qquad [P_i,N_j]=P_0 \delta_{ij} - H \epsilon_{ijk} J_k ~, \qquad [P_i,J_j]=-\epsilon_{ijk}
P_k ~,\nn\\
&&[N_i,N_j]=-\epsilon_{ijk} J_k ~, \qquad [J_i,J_j]=\epsilon_{ijk} J_k ~,~~~ [N_i,J_j]=-\epsilon_{ijk}
N_k ~, \label{clA}
\eea
with the first Casimir operator given by
\bea
\mathcal{C} = P_0^2-\vec{P}^2 + H (  \vec{P}  \cdot  \vec{N} + \vec{N}\cdot   \vec{P} ) - H^2 \vec{J}^2  ~. \label{Cacla}
\eea
The conserved charges are scalars under general coordinate transformations, but
one can easily verify that upon introducing formally the commutation relations
\be
[p^{\mu},x^{\nu}]=g^{\mu\nu},  \label{tonno}
\ee
one then obtains a set of ``noncommuting charges" which closes the same $SO(4,1)$ Lie algebra
as the associated generators.

We are describing the isometries of dS spacetime in terms of a set of generators
which are ``natural"
when using comoving coordinates. Their
algebraic properties (commutators) can be viewed
 as properties of the ``comoving-coordinates symmetry algebra".
By a general coordinates transformation an associated  isomorphic realization of the symmetry algebra
is found.
A general coordinates transformation acts as a
rotation between two different realizations of the isometry algebra.

There exists a variety of perspectives in which the motion of a particle in General Relativity can be examined.
For our purposes it is useful to focus on an approach based on
the representation (\ref{Pierro}) of conserved charges. We intend to focus
on the motion of massless particles, which is directly connected with
the causal structure of the theory. In preparation for the type of analysis  described in the following section, in which we
consider quantum-spacetime issues, we focus on the
case of the 1+1D subalgebra of 3+1D dS algebra\footnote{The
symmetry algebras for the 3+1D, 2+1D and 1+1D cases of the classical dS space-time are all contained into one another as sub-algebras.}.

As one can indeed infer from the analogous of (\ref{clA}) holding for conserved charges, the 1+1D dS  algebra  of the charges is characterized by
the commutators
\bea
&&[\Pi_E,\Pi_p]=H\Pi_p \qquad\qquad [\Pi_E,\Pi_N]=\Pi_p-H\Pi_N\nn\\
&&[\Pi_p,\Pi_N]= \Pi_E, \label{algebraofpi}
\eea
and the first Casimir in terms of the conserved charges can be written as:
\bea
\mathcal{C} = \Pi_E^2 - \Pi_p^2 + H (\Pi_p \Pi_N +\Pi_N \Pi_p)\,. \label{cas}
\eea

The explicit expressions of the conserved charges are
\bea
&&\Pi_{E}=E+H x\, e^{2Ht} p , \label{pie}\\
&&\Pi_{p}=-e^{2Ht} p , \label{pip}\\
&&\Pi_{N}=x \Pi_E + \left(\frac{1-e^{-2Ht}}{2H}+\frac{H}{2}x^2 \right) \Pi_p .\label{pin}
\eea

The Casimir relation (\ref{Cacla}) for the conserved charges leads to
the dS mass-shell condition:
\be
m^2=E^2-e^{2Ht}{p}^2.\label{eq:massshell}
\ee
Note that, while $\Pi_E$ and $\Pi_P$  are the conserved quantities, the observable
one-particle energy $E$ and momentum $p$ are not conserved,
and in particular they scale as $E_1a_{dS}(t_1)=E_2 a_{dS}(t_2)$,
as one can indeed infer from (\ref{eq:massshell}), (\ref{pie}) and (\ref{pip})
 for a massless particle, consistently with the scaling
induced by cosmological redshift in a dS Universe.

Let  us consider now the 1+1D motion of a photon in 3+1D dS spacetime and derive the
expression for the distance travelled by the photon starting at time $-t_0$ and observed at time $t$.
Let us notice that if $x(-t_0)=0$, then
\bea
\Pi_N (\{x=0,t=-t_0\}) = \mathcal{N}\equiv\frac{1-e^{2 H t_0}}{2 H}\Pi_p ~. \label{pin=0}
\eea
For a massless particle the Casimir equation (\ref{cas})  takes the form
\begin{equation}
 \Pi_E=\pm\Pi_p \sqrt{1-\frac{2 H \mathcal N}{\Pi_p}}~,
\end{equation}
which can be rewritten as follows
\begin{equation}
 \Pi_E=-\Pi_p e^{ H t_0},\label{eq:PeVSPp}
\end{equation}
using the explicit expression (\ref{pin=0}) of $\mathcal N$ (and fixing the sign ambiguity by choosing
to consider a case in which $E=p$ would hold in the $H\rightarrow 0$ Minkowski limit).

Denoting again by $\mathcal N$ the value of the conserved $\Pi_N$ along geodesics, we can rewrite it as
\begin{equation}
 \mathcal N= -\Pi_p e^{ H t_0} x  + \left(\frac{1-e^{-2Ht}}{2H}+\frac{H}{2}x^2 \right) \Pi_p,
\end{equation}
where we have substituted $\Pi_E$ with its expression in  terms of $\Pi_p$, eq. (\ref{eq:PeVSPp}).
Solving the equation above for $x$ we find
\begin{equation}
 x_{dS}(t)=\frac{ e^{ H t_0}\pm e^{- H t} }{H},\label{eq:xdS}
\end{equation}
where only the minus sign is consistent with the initial condition $x(t=-t_0)=0$.

\section{q-dS and its $\kappa$-Poincar\'e/$\kappa$-Minkowski limit} \label{q-dS}

In order to provide an illustrative example of the possible interplay between
curvature and Planck-scale-induced quantum corrections we analyze a quantum description
of dS spacetime such that its $H \rightarrow 0$ limit provides a well-known
quantum description of Minkowski spacetime, the $\kappa$-Minkowski noncommutative
spacetime~\cite{majrue,kpoinap,gacmaj,lukieFT,gacmich,wesskappa}.
We find convenient to derive most results in terms of the properties of the
algebra of symmetries of the quantum spacetime, rather than on the dual~\cite{FRT}
spacetime-coordinate picture.
This is the approach which turned out to be most fruitful also
in the study of theories
in $\kappa$-Minkowski~\cite{majrue,kpoinap,gacmaj,lukieFT,gacmich,wesskappa}.
Just as $\kappa$-Minkowski could be described fully
as the noncommutative spacetime dual to the $\kappa$-Poincar\'e Hopf
algebra~\cite{lukieIW,majrue,kpoinap},
q-dS spacetime can be introduced as the spacetime dual to the q-dS Hopf algebra.

In the 3+1D q-dS case this spacetime/spacetime-symmetry picture is still only developed rather
poorly~\cite{Ballesteros:2004eu}.
We shall therefore base our intuition and obtain our results in the 2+1D and 1+1D dS cases.
One indeed  finds explicit formulations of the 2+1D and 1+1D q-dS Hopf algebras
in the literature \cite{lukieIW,lukieAM},
but we must stress that the relation between them is not simple,
as a result of the non-embedding property of the Drinfel'd-Jimbo
deformation of dS algebra.
Unlike in the case of their Lie-algebra limits (mentioned in the preceding section),
the 1+1D q-dS Hopf algebra cannot be obtained as a simple restriction of the 2+1D q-dS Hopf algebra,
and (in spite of the preliminary nature of the results so far available on the 3+1D case)
we of course expect that a similar complication affects the relationship between the 3+1D and
2+1D cases.
However, one can also see \cite{lukieIW,lukieAM}
that the differences between the 1+1D restriction of the 2+1D q-dS Hopf algebra
and the 1+1D algebra are not of a type that should lead to sharp changes in the physical picture
and it is natural to expect that, once an explicit formulation for the 3+1D case will be available,
the 3+1D case will also turn out to be rather similar to the other ones.
One should therefore be able
to obtain a rather reliable first look at q-dS theories by considering the 2+1D and even the 1+1D case.

In the next subsection we start by reviewing briefly some well-known properties
of the $\kappa$-Minkowski spacetime that are particularly significant for our analysis.
Then in Subsection \ref{sec:qdS2+1} we discuss some properties of the 2+1D q-dS Hopf algebra, focusing on the
aspects that are most relevant for our analysis of the interplay between curvature and
Planck-scale effects. A similar description of the 1+1D q-dS Hopf algebra is given in Subsection~\ref{sec:qdS},
and a few remarks on the q-dS spacetimes are offered in Subsection~\ref{asi}.

\subsection{Some key aspects of $\kappa$-Minkowski spacetime}
The $\kappa$-Minkowski noncommutative spacetime has
coordinates that satisfy the commutation relations~\cite{majrue,kpoinap,gacmaj,lukieFT,gacmich,wesskappa}
\bea
&[x_{0},x_{j}]=i\lambda x_{j} \nonumber \\
&[x_{j},x_{k}]=0,     \label{kappamin}
\eea
where the noncommutativity parameter\footnote{In most of the \kM\
literature one finds the equivalent parameter $\kappa$, which is
$\kappa = 1/\lambda$, but our formulas turn out to be more compact
when expressed in terms of $\lambda$.} $\lambda$ is often assumed to
be proportional to the Planck length scale.

Even just a quick look at the commutation relations (\ref{kappamin})
already suggests that, while space-rotation symmetry remains classical,
translation and boost symmetries are modified by the $\kappa$-Minkowski noncommutativity.
These modified symmetries, as well as other properties of theories in $\kappa$-Minkowski~\cite{aad},
are very naturally
described in terms of a ``Weyl map"~\cite{wess}, a one-to-one map between elements
of the space of
functions of the $\kappa$-Minkowski noncommutative coordinates and elements of
the ordinary space of functions of commuting coordinates.
 It is sufficient to specify such a Weyl map $\Omega$ on the
complex exponential functions and extend it to the generic
function $\phi(x)$, whose Fourier transform is
$\tilde{\phi}(k)=\frac{1}{(2\pi)^4}\int d^4x\,\phi(x)e^{-ikx}$, by
linearity~\cite{aad}.
For instance, using a time-to-the-right ordering convention,
\be
\Phi(\x)\equiv\Omega(\phi(x))=\int d^4k\,
\tilde{\phi}(k)\, \Omega(e^{ikx}) =\int d^4k \,\tilde{\phi}(k)\,
e^{-i \vec{k} {\cdot} \vec{\x}}e^{ik_0{\x}_0} \label{kMelem}\,.
\ee
(We are adopting conventions such
that $k x \equiv k_\mu x^\mu \equiv k_0 x^0 - \vec{k} {\cdot} \vec{\x}$.)

It is relatively straightforward~\cite{aad}
to see that, consistently with this choice of Weyl map,
the action of generators of translations, $P_{\mu}$, and space-rotations, $M_j$,
should be described as follows
\bea
&&P_{\mu}\Phi({\x})=\Omega[-i\partial_{\mu}\phi(x)]\,,  \label{bicrossbasisA} \\
&&M_j\Phi({\x})=\Omega[i\epsilon_{jkl}x_k\partial_l\phi(x)] ~. \label{bicrossbasisB}
\eea
This means that for both
translations and space-rotations
one can introduce a ``classical action" (classical through the Weyl map).
However, while rotations are truly classical, one can easily see that
(as one expects on the basis of the form of the $\kappa$-Minkowski
commutation relations) translations are not fully classical.
There is no deformation in the ``action rule" (\ref{bicrossbasisA}) of translations,
but a deformation necessarily appears in the ``Leibnitz rule", {\it i.e.} the
noncommutativity scale enters in the rule for the action of translations
on the product of functions of the noncommutative coordinates.
We can see this already by considering the implications of the action rule (\ref{bicrossbasisA})
for the action of translations on a product of two Fourier exponentials:
\bea
&&P_j\Omega(e^{ikx})\Omega(e^{ipx})
=-i\Omega(\partial_j e^{i(k \cplus p)x})=\nn\\
&&=-i\Omega((k\cplus p)_je^{i(k\cplus p)x})=\nn\\
&&=[P_j\Omega(e^{ikx})][\Omega(e^{ipx})]
+[e^{-\lambda P_0}\Omega(e^{ikx})][P_j\Omega(e^{ipx})]
~,
\label{calccoprod}
\eea
where $p \cplus q \equiv (p_0+q_0, p_1 + q_1 e^{-\lambda p_0},
 p_2 + q_2 e^{-\lambda p_0}, p_3 + q_3 e^{-\lambda p_0})$ characterizes the product of
 exponentials in just the correct way to reflect the noncommutativity of the spacetime
 coordinates on which those exponentials depend on.
 For this type of deformations of the Leibnitz rule one speaks of the presence
 of a ``nontrivial coproduct". For example in the case we are now considering,
 the coproduct of space translations $\Delta P_j$,
 one sees from (\ref{calccoprod}) that
\be
\Delta P_j=P_j\otimes 1
+ e^{-\lambda P_0}\otimes P_j ~. \label{coprodpr}
\ee
Following an analogous procedure one can verify that instead the coproduct of time translations
is trivial
\be
\Delta P_0=P_0\otimes 1
+ 1 \otimes P_0 ~. \label{coprodpt}
\ee
While for translations the deformation is only in the Leibnitz rule,
for boosts there is even a deformation of the action rule.
One finds
\bea
N_j\Phi({\x})=\Omega(-[ix_0\partial_j
+x_j(\frac{1-e^{2i\lambda \partial_0}}{2\lambda}
-\frac{\lambda}{2}\nabla^2) +\lambda
x_l\partial_l\partial_j]\phi(x)) ~. \label{bicrossboost}
\eea
This result can be derived in several independent ways. One possibility
is just to insist on the ``consistency of the Hopf algebra": one speaks of a
symmetry Hopf algebra when the commutators and coproducts of the symmetry generators
close on the generators themselves.
If one for example replaced (\ref{bicrossboost}) with the classical action of boosts
then the coproduct of boosts would require~\cite{aad} the introduction of operators
external to the algebra.

The generators introduced in (\ref{bicrossbasisA}), (\ref{bicrossbasisB}),
(\ref{bicrossboost}) close the well-known $\kappa$-Poincar\'e Hopf
algebra~\cite{lukieIW,majrue,kpoinap}. As it is always the case for a Hopf algebra,
different choices of generators for the algebra lead to formulations that are
apparently rather different, the so-called different ``bases" of the Hopf algebra.
In the case of the $\kappa$-Poincar\'e Hopf
algebra~\cite{lukieIW,majrue,kpoinap} the different bases have a simple description
in terms of the ordering conventions adopted on the dual $\kappa$-Minkowski side.
We have adopted (see (\ref{kMelem})) the
time-to-the-right convention,
which is preferred by most authors~\cite{majrue,gacmaj,lukieFT,majoec},
and, in order to avoid potential complications which are
unrelated to the point we are making, we will work throughout consistently with this choice
of conventions, even at the level of the generalization to the case of q-dS algebra and
spacetime.

For the generators introduced in (\ref{bicrossbasisA}), (\ref{bicrossbasisB}),
(\ref{bicrossboost}) one obtains the following $\kappa$-Poincar\'e
commutators
\bea
\qs P_{\mu},P_{\nu}\qd&=&0\,,\nn\\
\qs M_j,M_k\qd&=&i\varepsilon_{jkl}M_l,\;\;\;\qs {N}_j,M_k\qd
=i\varepsilon_{jkl}{N}_l,\;\;\;\qs {N}_j,{N}_k\qd
=-i\varepsilon_{jkl}M_l\,,\nn\\
\qs M_j,P_0\qd&=&0,\;\;\;[M_j,P_k]=i\epsilon_{jkl}P_l\,,\nn\\
\qs {N}_j,P_0\qd&=&iP_j\,,\nn\\
\qs {N}_j,P_k\qd&=&i\qs\ts \frac{1- e^{-2\lambda
P_0}}{2\lambda}+\frac{\lambda}{2}\vec{P}^2\td\delta_{jk} -\lambda
P_jP_k\qd\,,\nn
\eea
and coproducts
\bea
\Delta(P_0)&=&P_0\otimes 1+1\otimes P_0\,,\,\,\,\;\;\;\Delta(P_j)
=P_j\otimes 1+e^{-\lambda P_0}\otimes P_j\,,\nn\\
\Delta(M_j)&=&M_j\otimes 1+1\otimes M_j\,,\nn\\
\Delta({N}_j)&=&{N}_j\otimes 1+ e^{-\lambda P_0}\otimes
{N}_j-\lambda\epsilon_{jkl}P_k\otimes M_l ~.
\eea
 Correspondingly the ``mass-squared" Casimir operator, $C_{\lambda}$,
takes the form
\be
C_{\lambda} = \frac{1}{\lambda^2}\left[\cosh(\lambda m)-1 \right]=(2/\lambda)^2\sinh^2(\lambda P_0/2)-e^{\lambda
P_0}\vec{P}^2  ~, \label{casimKAPPAMIN}
\ee
where we also introduced the so called mass parameter $m$, which is expected to
describe the rest energy.

This Casimir relation has generated significant interest in the quantum-gravity/quantum-spacetime
literature. This interest originates mainly from
taking the working assumption that the mass-Casimir relation between the
generators $P_0,\vec{P}$ might reflect the form of the (dispersion) relation
between energy $E$ and momentum $p$ (expected to still be conserved charges
derivable from the presence of symmetry under $P_0,\vec{P}$).
If this working assumption is correct then there could be some striking effects,
including
a dependence of speed
on energy for massless particles, $v_{(m=0)} = e^{\lambda E}$
 (obtained from the dispersion relation using
 the familiar law $v=dE/dp$).
If $\lambda$ is of the order of the Planck length, such a velocity law would
fit naturally within a rather wide quantum-gravity literature which, for independent
reasons, has been considering analogous laws~\cite{grbgac,gampul,mexweave,leereview}.
At energies accessible in laboratory experiments one can always safely
assume $e^{\lambda E} \simeq 1$, but in the early stages of evolution of the Universe
the typical particle energy was extremely high, and some authors have discussed~\cite{infla1,maguREV}
the possibility that such laws of energy dependence of the speed of massless particles
might have significant implications for our understanding of the early Universe,
with significance both for inflation and possibly other features that
are relevant for establishing which regions of the
Universe were in causal connection at a certain era in the evolution.

Recent results~\cite{kappa1,kowafreid,kappa2,kappa3,theta,ArMa1,ArMa2}
suggest that the assumption that the energy-momentum relation should exactly reproduce
the Casimir relation between symmetry generators might have to be improved upon\footnote{The
studies reported in Refs.~\cite{kappa1,kowafreid,kappa2,kappa3,theta,ArMa1,ArMa2}
showed that the Noether technique of derivation of conserved charges,
and in particular of the energy/momentum charges associated to space/time translational invariance,
are applicable also to the case of field theories with Hopf-algebra spacetime symmetries
formulated in noncommutative spacetimes. There are still some challenges concerning the
(``operative") interpretation of the charges that are derived in these novel Noether
analyses, but the preliminary indications that are emerging suggest that
the relation between energy-momentum charges might be somewhat different
from the (Casimir) relation between the translation generators with differences
that however do not change the nature (order of magnitude and energy-momentum dependence)
of the Planck-scale-induced correction terms. It is therefore still legitimate
to perform preliminary investigation of the implications of spacetime noncommutativity
assuming that the relation between charges roughly resembles the Casimir relation
for the symmetry generators.},
but also confirm that more careful analyses do not change significantly
the key expectations. This provides partial encouragement for us (see later)
to assume that a similar working assumption for the q-dS case can be reliably used
for a first preliminary level of investigation.

Most of our more quantitative results for q-dS case will focus on the 1+1D case,
so we close this subsection by noting
the commutators and coproducts for
the 1+1D $\kappa$-Poincar\'e Hopf algebra (written consistently with our time-to-the-right
conventions):

\begin{eqnarray}
&\left[ P_0,P\right] = 0, \qquad \left[N , P_0\right] = i P, \qquad \left[ N , P\right] = i  \frac{1}{2 \lambda}\left(1 - e^{-2\lambda
P_0}\right) - i\frac{\lambda}{2} P^2, \\
&\Delta(P_0) = P_0 \otimes 1 +  1 \otimes P_0~, \!\!\!\!\qquad \Delta(P) =  P \otimes 1 + e^{-\lambda P_0} \otimes P~, \qquad \!\!\!\!\! \Delta (N)
= N \otimes  1 + e^{-\lambda P_0} \otimes N~.\nonumber
\end{eqnarray}

\subsection{q-dS algebra of symmetries in 2+1D}  \label{sec:qdS2+1}
As mentioned above, we shall mainly consider
 q-dS algebras for  2+1D and 1+1D cases, consistently adopting throughout
 conventions such that these q-dS Hopf algebras contract (in the In\"on\"u-Wigner sense)
 to the formulation of the $\kappa$-Poincar\'e algebra
 given in the conventional ``time-to-the-right basis".
We shall further specify our conventions by demanding that the ``classical limit"
of our basis for q-dS reproduces the classical dS
algebra written for comoving coordinates.

Let us start by noting down the commutators and coproducts which characterize
our description of the q-dS Hopf algebra in the 2+1D case. The commutators are
\bea
&[J,P_0]=0,\qquad [J,P_i]=\varepsilon_{ij} P_j,\qquad [J,N_i]=\varepsilon_{ij} N_j,\nonumber\\
&[P_0,P_i]= H P_i,\qquad [P_0,N_i]=P_i-H N_i, \qquad [P_1,P_2]=0 \nonumber \\
&[P_i,N_j]=-\delta_{ij}\left (H \f{e^{- {2 w \f{P_0}{H}  }    }}{2 w}
-H\f{\cos\left ( 2 w J\right )}{2 w} + \f{1}{2}\tanh\left ( {w} \right )\left
( \{P_i,N_i\}- \f{\vec{P}^2}{H}\right )\right )+ \nonumber \\
&-\tanh \left({w} \right )\left ( \f{P_jP_i}{H}+\varepsilon_{ij}\f{H}{2
w} \sin(2 w J)  -(P_jN_i+N_jP_i)\right )-\varepsilon_{ij}\f{H}{2 w} \sin(2 w J) ,\nonumber \\
&[N_1,N_2]= - \f{1}{2 w} \sin(2 w J)~,
\eea
where $\varepsilon_{ij}$ is the Levi-Civita tensor ($i,j \in \{1,2\}$;
$\varepsilon_{12}=1$) and we used notation consistent with the one introduced in the previous
section for the classical limit (the $w \rightarrow 0$ ``classical" limit of our description of
the q-dS Hopf algebra reproduces the description of the dS Lie algebra given in the previous section).

The coproducts are
\bea
&\Delta (P_0)=1\otimes P_0 + P_0\otimes 1, \qquad \Delta (J)=1\otimes J + J \otimes 1,\nonumber\\
&\Delta (P_i)= e^{- { w \f{P_0}{H}  }    }\otimes P_i + P_i\otimes \cos(w J)-\varepsilon_{ij}
P_{j}\otimes \sin(w J),\nonumber\\
&\Delta (N_i)= e^{- { w \f{P_0}{H}  }    }\otimes N_i + N_i\otimes \cos(w J)
+\varepsilon_{ij} \f{P_{j}}{H}\otimes {\sin(w J)} ~,
\eea
and the q-dS first Casimir is given by
\bea
&\mathcal{C}=4H^2 \cosh(w)\left [  \f{\sinh^2\left ( \f{w P_0}{2H}\right
)}{w^2}\cos^2\left ( \f{w J}{2}\right ) -  \f{\sin^2\left (
\f{w J}{2}\right )}{w^2}\cosh^2\left ( \f{w P_0}{2  H}\right ) \right
]- \f{\sinh(w)}{w}e^{w \f{P_0}{H}}  \cdot \nonumber \\
& \cdot   \left[ \cos(w J)(\vec{P}^2-H\{N_i,P_i\})+
2 H \sin( w J)\left(( P_1 N_2-P_2 N_1)+ H\f{1}{2w}\sin(2wJ)\right) \right]~. \label{Casimir}
\eea

This casimir relation will play a key role in our analysis.
We shall analyze it mainly for what concerns the implications it suggests
for conserved charges, in the spirit of the observations reported
at the end of the previous subsection.
Specifically we shall assume that, also in the q-dS case,
 the charges
satisfy the same algebraic relations as the generators once the commutation relations (\ref{tonno})
are formally introduced.
Using this reasoning in reverse one can estimate the properties of the charges by
looking at the ones of the generators, and taking into account the implications
of introducing formally (\ref{tonno}).
We should warn our readers that in the dS case, besides the limitations of this approach
already debated in the quantum-Minkowski literature (here briefly mentioned in the preceding
subsection), there are additional challenges which originate in some ``ordering issues".
In particular, for $H \neq 0$ 
the relevant Casimir acquires a dependence on noncommuting generators.
For $H=0$ the Casimir depends only on $P_0$
and $P_i$, and they commute for $H=0$. But for $H \neq 0$ one finds that the generators $P_0$
and $P_i$ do not commute and in addition the Casimir also acquires a dependence on the boost
generators $N_i$ which of course does not commute with $P_0$
and $P_i$. This results in an ambiguity
for the implementation of the ``recipe" of substitution of generators by numerical values
of charges carried by a classical particle. We shall not dwell much on this ``ordering issue"
for $P_0$ and $P_i$ and present
results adopting only one particular (and not necessarily compelling) choice or ordering.
Some quantitative details of the formulas we produce do depend on this ordering ambiguity,
but for the qualitative features we do highlight, which are the main objective of our analysis,
 we have verified that they are robust under changes of ordering convention.

Concerning the In\"on\"u-Wigner contraction, which classically takes dS to Minkowski/Poincar\'e,
it is rather significant for our analysis
that one can find both contractions of q-dS to $\kappa$-Minkowski/$\kappa$-Poincar\'e
and contractions of q-dS to classical Minkowski/Poincar\'e.
The outcome of the In\"on\"u-Wigner contraction procedure
depends crucially on the relationship
between $H$ and the quantum-group deformation parameter $w$.
We can easily show this feature since
we have already described the algebras in terms of
appropriately ``$H$-rescaled generators"~\cite{lukieIW,inonuwigneroriginal},
and therefore the contraction will be achieved at this point by simply
taking $H \rightarrow 0$.

For small values of $H$ some quantum-gravity arguments
(see, {\it e.g.}, Ref.~\cite{kodadsr} and references
therein) suggest that
the relation between $H$ and $w$ should be well approximated,
for small $H$,
by a parametrization in terms of a single parameter $\alpha$:
\be
w \sim \left( H L_p \right)^{\alpha} \label{ant} ~,
\ee
where $L_p$ is the Planck length scale ($\simeq 10^{-33}cm$) and the parameter $\alpha$
may depend on the choice of quantum-gravity model.
By inspection of the formulas given above one easily finds that,
depending on the value of this parameter $\alpha$, the $H \rightarrow 0$ contraction of the
q-dS Hopf algebra for 2+1D spacetime leads to the following possible results:
\begin{itemize}
\item If $\alpha=1$ the
contraction of the 2+1D q-dS Hopf algebra gives the 2+1D $\kappa$-Poincar\'e Hopf algebra.
In particular, for $\alpha=1$ and small $H$ one finds that the ``q-dS mass Casimir"
takes the form
\bea
C \big|_{small~H} \simeq \f{4}{L_p^2} \, \sinh^2 \left( \f{L_p P_0}{2} \right)
- e^{L_p P_0}\vec{P}^2
+O(H) ~, \label{jocZ1}
\eea
which is clearly consistent with the $\kappa$-Poincar\'e mass Casimir (\ref{casimKAPPAMIN}).
This result (\ref{jocZ1}) provides an example of the case in which,
at the level of infinitesimal symmetry transformations, quantum-spacetime corrections
for small values of the curvature are
curvature independent ($H$-independent).
But even for these cases where $w \simeq H L_p$ there is room for
significant (see below) source of interplay between curvature and Planck-scale effects,
originating from the fact that the quantum-gravity literature invites one to
contemplate different case of the relationship between $w$, $H$ and $L_p$
with the common feature of taking the shape $w \simeq H L_p$ in (and only in)
the small-$H$ limit.
\item  If $1 < \alpha < 2$ the $H \rightarrow 0$ contraction
of the 2+1D q-dS Hopf algebra gives the 2+1D {\underline{classical}} Poincar\'e (Lie) algebra,
and for small $H$
one finds that the ``q-dS mass Casimir"
takes the form
\be
C \big|_{small~H} \simeq P_0^2 - \vec{P}^2 - L_p (H L_p)^{\alpha -1} P_0 \vec{P}^2 + O(H) ~.
\ee
This case $1 < \alpha < 2$ clearly is an example of very strong interplay
between Planck scale and curvature, even for small curvatures (small values of $H$).
So much so that when $H=0$ there are no quantum-spacetime effects (at the symmetry-algebra level)
whereas as soon as $H \neq 0$ one finds the Planck-scale corrections.
This provides very clear evidence in support of our thesis: the interplay between curvature and Planck-scale
effects may be very significant, and even for small values of curvature.
\item If $\alpha \geq 2$ one stills finds (as in the case $1 < \alpha < 2$) that the contraction
of the 2+1D q-dS Hopf algebra gives the 2+1D classical Poincar\'e (Lie) algebra.
But with these high values of $\alpha$ one may say that
the quantum-algebra corrections are negligible even for small but nonzero values of
the curvature. In particular, for $\alpha \geq 2$ and small $H$
the ``q-dS mass Casimir"
takes the form
\be
C \big|_{small~H} \simeq P_0^2 - \vec{P}^2  ~.
\ee
\item Finally the case $\alpha < 1$ must be excluded since it provides an inconsistent
description of the Minkowski limit:
for $\alpha < 1$ the $H \rightarrow 0$ contraction of the 2+1D q-dS Hopf algebra
is affected by inadmissible divergences. Indeed there is no known
example~\cite{kodadsr} of a quantum-gravity argument
favouring a relationship between $w$, $H$ and $L_p$ characterized by $\alpha < 1$.
\end{itemize}

These observations show that the interplay between curvature and Planck-scale
(quantum-spacetime) effects can be very significant also in the small-curvature
limit. And we shall argue that this interplay can be even more significant
in the large-curvature limit.
Of course, in order to describe the behaviour for large values of
curvature we cannot rely on (\ref{ant}), which is considered in the quantum-gravity
literature only as a good approximation scheme for the small-curvature case.
And of course the complexity of quantum-gravity theories
represents a huge challenge
for attempts to estimate nonperturbative features, such as the exact form of the relation
between $H$ and $w$.
For our purposes it is however useful to adopt even a tentative {\it ansatz}
 for the exact form of the relation
between $H$ and $w$, since it allows us to give definite formulas that illustrate
the implications of curvature for quantum-spacetime effects very explicitly,
particularly by exposing differences between
the small-curvature and the high-curvature regimes.
With these objectives in mind we can consider the possibility
\be
w=\frac{2\pi}{2+\frac{1}{H L_{p}}} ~.
\label{wlqg}
\ee
which is inspired\footnote{We are
prudently stressing that Eq.~(\ref{wlqg}) for our analysis
is only loosely inspired by previous results in the quantum gravity literature
because of awareness of several subtleties that should be properly investigated
before establishing the relevance more robustly.
It is clearly encouraging for our study and for all the q-dS-based quantum-gravity
research to notice that the introduction of the cosmological
constant in $2+1$D canonical quantum gravity allows the resulting gauge
symmetry of the theory to be described in terms of
Hopf algebras/quantum groups~\cite{MaSmo, smolo}.
So the quantum groups symmetries are, in an appropriate sense, not an {\it a priori}
choice of hypothesis for these models, but rather something that is constructively
derived.
However, for what concerns specifically
 the relation between $w$ and $H$ codified in Eq.~(\ref{wlqg})
 we should stress that this is found~\cite{BNR} specifically
 in the study of $2+1$D gravity
 with negative cosmological constant rewritten as a Chern-Simons theory.}
 by results reported in the literature
on $2+1$D canonical quantum gravity~\cite{MaSmo, smolo,strabo, BNR}.

The formula (\ref{wlqg}) is intriguing from our perspective
since it reduces to $w \simeq H L_p $ ({\it i.e.} the case considered above with $\alpha =1$)
for $H$ much smaller than $\frac{1}{L_p}$,
but then for large values of $H$ the quantum-algebra deformation
becomes essentially constant
(in the sense that $w \simeq \pi$) and independent of $L_p$.
This is therefore an example in which the case of large curvatures eliminates from our
theoretical framework any dependence on the Planck scale, even though the Planck-scale
effects are very significant in the small-curvature regime.

\subsection{q-dS algebra of symmetries and charges in 1+1D } \label{sec:qdS}
It is easy to verify that the same type of interplay between curvature and quantum-geometry/quantum-algebra
effects discussed in the previous subsection for the case of the 2+1D q-dS Hopf algebra
is also found in the case of the 1+1D q-dS Hopf algebra.
But we nevertheless find appropriate to report a few observations on the
1+1D q-dS Hopf algebra, since this will provide the basis for a tentative analysis
of a ``quantum FRW spacetime" proposed in Section \ref{sec:qFRW}.

Let us start by noting down some key characteristics of
the 1+1D q-dS Hopf algebra~\cite{vulpi}, adopting conventions for the choice of ``basis"
that are consistent with the corresponding ones adopted in the previous subsections.
The commutators are
\bea
\!\!\!\!\!\!\!\!\!\!\!\!\!\!\!\!\!\!\!\!\!\!\!\!\!\!\!\!\!\!\!\!\!\!\![ P_0,  P] &=&
HP, \qquad [ P_0,  N] = P-H N,\nn\\
\!\!\!\!\!\!\!\!\!\!\!\!\!\!\!\!\!\!\!\!\!\!\!\!\!\!\!\!\!\!\!\!\!\!\!{[} P,  N] &=& \cosh(w/2)\f{1-e^{\f{-2
w P_0}{H}}}{2w/H} - \f{1}{H}\sinh(w/2) e^{\f{-wP_0}{H}}\Theta
~, \label{1+1A2}
\eea
where we introduced, for compactness, the notation
$$
\Theta  = \qs e^{\f{wP_0}{2H}}  (P-H N) e^{\f{wP_0}{2H}}  (P-HN) - H^2
e^{\f{wP_0}{2H}}  N e^{\f{wP_0}{2H}} N \qd
~.
$$
For the coproducts one finds
\bea
\Delta(  P_0)=1\otimes   P_0 +   P_0\otimes 1,
\qquad& \Delta(  P)= { e}^{\f{-w   P_0}{H}}   \otimes   P
+  P\otimes  1, \qquad \Delta( N)= { e}^{\f{-w   P_0}{H}}\otimes  N +  N\otimes  1 ~, ~~~~~~~~&
\label{jocoprod}
\eea
and the mass Casimir is
\be
\mathcal{C}=  H^2\f{\cosh(w/2)}{w^2/4}\,{\sinh^2\ts\f{wP_0}{2H}\td}
- \frac{\sinh (w /2)}{w/2}\Theta \label{1+1C2} ~.
\ee

In Section \ref{sec:qFRW} we examine some properties of the propagation of massless particles
in a quantum FRW spacetime on the basis of some corresponding properties of massless particles
in 1+1D q-dS spacetime. We are therefore primarily interested in analyzing (\ref{1+1A2}),(\ref{1+1C2})
for the case of a massless particle.
In particular, from the form of the Casimir we infer that
for massless particles
\be
0 =  H^2  \f{\cosh(w/2)}{w^2/4}\,{\sinh^2\ts\f{wP_0}{2H}\td}  - \frac{\sinh (w /2)}{w/2}\Theta\,,
\label{casimirzero}
\ee
and therefore
\be
\sinh(w/2) \Theta = H^2 \f{\cosh (w/2)}{w/2} {\sinh^2\ts\f{wP_0}{2H}\td}\,.
\ee
This last equation can be used to obtain a simplified form
of the last commutator in (\ref{1+1A2}):
\be
[ P,N ] = \f{H}{w} \cosh(w/2) \ts  e^{\f{-w P_0}{H}}-  e^{\f{-2w P_0}{H}}     \td \,. \label{ncomm}
\ee
As discussed earlier, these properties at the algebra level can be used to
motivate some corresponding proposals for the conserved charges,
also using formally the commutation relations $[p^\mu,x^\nu]=g^{\mu \nu}$.
This leads us to the following representation of the charges\footnote{In order to verify
that the charges described in (\ref{dpip}) and (\ref{dpin}) (once formally endowed with
the noncommutativity implied by the commutation relations $[p^\mu,x^\nu]=g^{\mu \nu}$)
close the q-dS Hopf algebra it is sufficient to make use of the Sophus-Lie expansion
and of the
observation $[A,\f 1B] = -\f 1B [A,B] \f 1B$,
which is valid for any pair of noncommuting operators $A$ and $B$.}:
\bea
\!\!\!\!\!\!\!\!\!\!\!\!\!\!\!\!\!\!\!\!\!\!\!\!\!\!\!\!\!\Pi_E= E + He^{2Ht} x p\,, \qquad\qquad \Pi_p
=\partial_x = - e^{2Ht} p\,, \label{dpip}
\eea
\bea
\!\!\!\!\! \Pi_N \!\!\!  &=&\!\!\! \f{H }{2 w}\f{\cosh(w/2)}{ \sinh\! w}\, e^{-\f{w\Pi_E}{H}} \f{1}{\Pi_P}\,
e^{-\f{w\Pi_E}{H}}  \left( e^{-2w \Pi_P x}-1 \right )
+  \nn\\ \!\!\!&-&\!\!\! \f{H }{2 w}\f{1}{ \tanh(w/2)}\, e^{-\f{w\Pi_E}{2
H}} \f{1}{\Pi_P} \, e^{-\f{w\Pi_E}{2H}}  \left ( e^{-w \Pi_P x}
-1 \right )    + \f{1-e^{-2Ht}}{2H} \Pi_P ~.\label{dpin}
\eea
Notice that the representations of $\Pi_E$, $\Pi_p$ are unchanged with respect to the corresponding
classical-spacetime case. The quantum-algebra deformation only affects the representation
of  $\Pi_N$. Of course, also $\Pi_N$ reduces to its classical-spacetime limit
upon setting $w\to 0$
\be
\lim_{w\rightarrow 0} \Pi_N = \f{1}{2} Hx^2 \Pi_p + x \Pi_E + \f{1-e^{-2Ht}}{2H} \Pi_P\,.
\ee
As discussed in Section~\ref{sec:dS}, the $\Pi_N$ charge plays a key role in
the derivation of the path of a massless particle, and therefore
one should expect that the quantum-algebra deformation of the
representation of the $\Pi_N$ charge affects significantly the description
of the path of a massless particle.
As in Section~\ref{sec:dS}  we observe that
\be
\mathcal N \equiv \Pi_N (\{x=0,t=-t_0\}) =\f{1-e^{2Ht_0}}{2H} \Pi_P
~.
\label{defconstrjo}
\ee
And enforcing this constraint (\ref{defconstrjo}) into the Casimir relation one obtains
\begin{equation}
 0=H^2\f{\cosh(w/2)}{w^2/4}\,{\sinh^2\ts\f{w \Pi_E}{2H}\td}- \frac{\sinh (w /2)}{w/2}\Theta_{\mathcal N},
\end{equation}
where $\Theta_{\mathcal N}  = \qs e^{\f{w\Pi_E}{H}}  (\Pi_P-H \mathcal N)^2  - H^2
e^{\f{w\Pi_E}{H}}  \mathcal N^2  \qd$.

Solving the above equation with respect to the variable $Y \equiv e^{-\frac{w \Pi_E}{H} }$
one finds two solutions:
\begin{equation}
Y_{\pm}=1\pm \sqrt{\frac{2w}{H^2}\tanh (w /2) \left[  (\Pi_P-H \mathcal N)^2  - H^2
 \mathcal N^2\right] },
\end{equation}
which, using (\ref{defconstrjo}), can be rewritten in the form
\begin{equation}
Y_\pm=1\pm \frac{ e^{Ht_0} \Pi_P }{H} \sqrt{2w\tanh (w /2)  } ~. \label{eq:y}
\end{equation}
We shall only take into account the solution $Y_+$, ignoring $Y_-$, since we are looking
for $\Pi_P=-\Pi_E$  in the  $w\rightarrow 0, H \rightarrow 0$ limit (see Sec. \ref{sec:dS}).
In the following we denote simply with $Y$ the solution $Y_+$.

Since $\Pi_N$ is conserved during the particle motion, we require it to be always equal to $\mathcal N$.
From the charge definition (\ref{dpin}) one then obtains
\begin{equation}
 \mathcal N=\f{H }{2 w}\f{1}{2 \sinh\! \frac{w}{2}}\,  \f{Y^2}{\Pi_P}\,
\left( e^{-2w \Pi_P x}-1 \right )
   -\f{H }{2 w}\f{1}{ \tanh(w/2)}\,\f{Y}{\Pi_P} \, \left ( e^{-w \Pi_P x}
-1 \right )    + \f{1-e^{-2Ht}}{2H} \Pi_P ~,
\end{equation}
and solving with respect to $e^{-w \Pi_P x}$ one finds
\begin{eqnarray}
e^{-w \Pi_P x} &=&\cosh{\frac{w}{2}}Y^{-1} -  \sqrt {\left(1-\cosh{(\frac{w}{2})}Y^{-1}\right)^2+2\frac{w}{H} \sinh{(\frac{w}{2})}\Pi_P Y^{-2}\left(-\frac{1-e^{-2Ht}}{H}\Pi_P +2 \mathcal N\right)}\nonumber \\
&=&\cosh{\frac{w}{2}}Y^{-1} -  \sqrt {\left(1-\cosh{(\frac{w}{2})}Y^{-1}\right)^2+2w \sinh{(\frac{w}{2})}\Pi_P^2 Y^{-2}\f{e^{-2Ht}-e^{2Ht_0}}{H^2} } ~,
\end{eqnarray}
where we also used (\ref{defconstrjo}) to eliminate $\mathcal N$,
and we removed a sign ambiguity by enforcing consistency with
the initial condition $x(t=-t_0)=0$.

So the deformed comoving distance travelled by a q-dS  mass-less particle,
that starts moving at time $t=-t_0$ is:
\begin{equation}
 x_{q-dS}(t)=-\frac{1}{w \Pi_P} \ln{\left[\cosh{\frac{w}{2}}Y^{-1}
 -  \sqrt {\left(1-\cosh{(\frac{w}{2})}Y^{-1}\right)^2+2w \sinh{(\frac{w}{2})}\Pi_P^2
 Y^{-2}\f{e^{-2Ht}-e^{2Ht_0}}{H^2} }\right]}\,.\label{eq:xqdS}
\end{equation}
This formula has the correct $w\rightarrow 0 $ limit, since
in this limit it reduces to the comoving distance travelled
by a massless particle  in dS spacetime (see eq.~(\ref{eq:xdS}))
\be
\lim_{w \to 0}  x_{q-dS} = \f{e^{H t_0}-e^{-Ht}}{H}  +O(w) ~.
\ee

Since $Y$ denotes the $Y_+$ of Eq.~(\ref{eq:y}) (and therefore $Y$ depends only
on $\Pi_P,w,H$),
our result (\ref{eq:xqdS}) gives the dependence of  $x_{q-dS}(t)$ on $\Pi_P,w,H$ and $t$.
Of course, if preferred, one can also use Eq.~(\ref{eq:y}) to trade the dependence
on $\Pi_P$ for a dependence on $\Pi_E$, obtaining
\begin{equation}
x_{q-dS}(t)  =\frac{\sqrt{2 w \tanh{(\frac{w}{2})}}}{w H e^{- H t_0}
(1-e^{\frac{-w\Pi_E}{H}})} \ln{[Z] } \label{eq:xqdSpe}
\end{equation}
with $Z=  \cosh{ (\frac{w}{2}) }  e^{\frac{w \Pi_E}{H}}
-\sqrt{  (1-\cosh{(\frac{w}{2})e^{\frac{w \Pi_E}{H}})^2
+\cosh{(\frac{w}{2})} e^{-2 H t_0}( 1-e^{ \frac{-w\Pi_E}{H} }  )^2  (e^{-2 H t}-e^{2 H t_0}) }  } $.

\subsection{Aside on quantum dS space-time} \label{asi}
We structure our analysis in such a way that
we can rely exclusively on the structure of the q-dS Hopf algebra of symmetries,
without any explicit use of the noncommutativity of the q-dS spacetime.
This, as mentioned, is consistent with an approach that has proven fruitful in the analysis
of other spacetimes that are dual to a Hopf algebra, such as $\kappa$-Minkowski.
Still some readers may find more intuitive a characterization of the spacetime
which is not only implicitly given in terms of a duality.
In closing this section we therefore provide an explicit description of the q-dS
noncommutative spacetime, relying on results previously obtained
in the literature~\cite{quantumgroup}.

The simplest strategy for obtaining an explicit description of the properties of the q-dS
spacetime coordinates uses  a procedure that performs a
semiclassical quantization
of the Poisson-Lie brackets \cite{quantumgroup},
based on the familiar Weyl
substitution~\cite{Drlb,Tak} of the Poisson brackets between
commutative coordinates by commutators between non-commutative coordinates.
While one can have quantum groups that do not
coincide with the Weyl quantization of its underlying Poisson-Lie
brackets, this procedure has proven fruitful in several previous applications
(see, {\it e.g.}, Refs.~\cite{LRZ,BBH} and references therein).

In Ref.~\cite{quantumgroup} one finds an explicit description of
the Poisson-Lie brackets for the 2+1D dS algebra\footnote{Notice that
the asymmetric form of the brackets with respect
to exchanges of coordinates is not an intrinsic feature of the
theoretical framework but rather a result~\cite{quantumgroup}
of the particular choice of the local coordinates $x_\mu$.}:
\be
\begin{array}{l} \displaystyle{ \{x_0,x_1\} =-w\,\frac{\tan H
x_1}{H^2 \cos^2 H x_2}, \qquad\{x_0,x_2\} =-w\,\frac{\tanh H x_2}{H^2 }, \qquad \{x_1,x_2\} =0 . }
\end{array}\nonumber \ee
From these one obtains the commutation rules for the coordinates of the 2+1D q-dS spacetime
 \be
 \begin{array}{l} {\displaystyle{ [\hat x_0, \hat x_1]
=-w\,\frac{\tan H \hat x_1}{H^2\cos^2
H \hat x_2}+o(w^2)= }} -\f{w}{H} \hat x_1 - \frac{1}{3} w H \hat x_1^3
- w H \hat x_1 \hat x_2^2 +o(w^2,H^2), \\[10pt]{\displaystyle{[\hat x_0,\hat x_2]
=-w\,\frac{\tan H \hat x_2}{H^2}+o(w^2)=}}   -\f{w}{H} \hat x_2-\frac 13 w H \hat x_2^3+o(w^2,H^2), \\[10pt]
[\hat x_1,\hat x_2] =0+o(w^2) .
\end{array} \label{ha}
\ee
As mentioned the q-dS spacetime reduces to $\kappa$-Minkowski
when an appropriate $H\to 0$ limit is taken. Indeed if one assumes
in (\ref{ha}) that for small $H$ the quantization parameter $w$ is proportional
to $H$, {\it i.e.} $w \simeq \lambda H$ for some $\lambda$,
then the $H\to 0$ limit of the commutation relations (\ref{ha})
reproduces the commutation relations of the $\kappa$-Minkowski
spacetime coordinates.

One may also introduce a description of the q-dS spacetime in terms
of non-commutative ambient (Weierstra\ss)
coordinates $(\hat \s_3,\hat \s_\mu)$, which reads~\cite{quantumgroup}
\be
\begin{array}{l} [\hat \s_0, \hat  \s_i]
=-\f{w}{H}   \,\hat  \s_3\hat  \s_i+o(w^2), \qquad [\hat  \s_1,\hat  \s_2]
=0+o(w^2)\,, \\[2pt]  [\hat  \s_3,\hat \s_0]=-w H\,\hat \s^2+o(w^2), \qquad\ \
   [\hat \s_3,\hat \s_i]=-w H \, \hat \s_0\hat \s_i +o(w^2)\,. \end{array} \label{hb}
   \ee
In this formulation the  symmetry under exchange of $\hat \s_1$ and $\hat \s_2$
is manifest, and, since $\hat \s_3\to 1 $ when $H\to 0$,
the first two relations   in (\ref{hb}) are directly connected to
the corresponding properties of the \kM $~$coordinates.

\subsection{Some possible applications of the q-dS algebra}\label{jocperspe}
The research effort we report in this manuscript was aimed at establishing
as robustly as possible the significance of the interplay between curvature
and Planck-scale effects, thereby correcting a commonly-adopted assumption
in quantum-gravity-phenomenology research.
In a certain sense the q-dS formalism is viewed within our analysis
as a toy model that is well suited for exposing fully our concerns
that it is not legitimate to
assume absence of
interplay between curvature and Planck-scale effects.
In closing this Section on the q-dS formalization we find appropriate to stress
that we feel that this formalization may well deserve more interest than
the one from the toy-model perspective, although we shall not impose this
intuition on our readers elsewhere in the manuscript.

One of the reasons for our choice to focus nearly exclusively
on the significance of the interplay between curvature and Planck-scale effects
is that this aspect has emerged from our investigations as a fully
robust feature,
qualitatively independent of the choices of perspective and ordering conventions we adopted.
Up to relatively uninteresting quantitative details the same significance of interplay
between curvature and Planck-scale effects is easily found even adopting
choices of ordering convetion that are different from the one on which
we focused for simplicity. And similarly one finds exactly the same level
of interplay between curvature and Planck-scale effects upon changing
the basis for the q-dS Hopf algebra, going for example from the one we
here preferred (because of its relevance for the much-studied ``time-to-the-right basis"
of the $\kappa$-Poincar\'e algebra)
to one obtained even by nonlinear redefinition of the generators.

This robustness of the significance of the
interplay between curvature and Planck-scale should have profound implications
for the directions to be taken in parts of the quantum-gravity-phenomenology literature,
but it is of course not at all surprising within the framework we adopted. As we stressed
already in the opening remarks
of this manuscript, the striking feature of the q-dS framework is that the key novel
structures depend on a single parameter $w$ which by construction is dimensionless.
So the only opportunities
for the Planck scale to appear in the description of the structure
of our quantum spacetime necessarily involve expressing this dimensionless
parameter in terms of the Planck scale and of the only other dimensionful scale
present in the framework, which is indeed the curvature scalar.

Of course it also interesting to examine the q-dS framework looking for features
that are of interest even beyond the issue of establishing the presence of a strong interplay between
curvature and Planck-scale effects. Our perception is that these specific features
might be more sensitive to possible ``changes of Hopf-algebra basis" and
possible alternative ways to handle the ordering issues discussed above.
But they are nonetheless interesting and we want to comment on at least a couple of them.

Probably the most significant of these features concerns the possibility
of describing a ``minimum-wavelength principle" in a framework that allows for
curvature.
The idea of
a ``minimum-wavelength principle" is justifiably popular in the quantum-gravity literature
and in fact several flat-spacetime formalizations have been proposed,
but to our knowledge the framework we developed here is the first example
of a possible description of a ``minimum-wavelength principle" in presence of curvature.
One way to see this is based on our equation (\ref{eq:y}) which (since we worked
with $Y \equiv e^{-\frac{w \Pi_E}{H} }$ and $\Pi_p = - |\Pi_p|$) establishes that
\begin{equation}
 e^{-\frac{w \Pi_E}{H} } = 1 -  \frac{ e^{Ht_0} |\Pi_P| }{H} \sqrt{2w\tanh (w /2)  } ~.
 \label{jocextra}
\end{equation}
Let us assume for definiteness that $w = H L_p$ and let us
first notice that this equation produces a ``minimum-wavelength principle"
in the $H \rightarrow 0$ limit, which is indeed the ``minimum-wavelength principle"
that motivated a significant portion of the interest in the $\kappa$-Minkowski/$\kappa$-Poincar\'e
framework:
\begin{equation}
 e^{-L_p \Pi_E } = 1 - L_p |\Pi_P|~.
\end{equation}
This indeed reflects the known mechanism for exposing a minimum wavelength (maximum $|\Pi_P|$)
in in the $\kappa$-Minkowski/$\kappa$-Poincar\'e
framework: the maximum allowed value for $|\Pi_P|$ is $|\Pi_P| = 1/L_p$,
as $|\Pi_P|$ approaches the value $1/L_p$ the flat-spacetime energy $\Pi_E$ diverges,
and for hypothetical values of $|\Pi_P|$ greater than $1/L_p$ there is no real-energy solution.
For values of $H$ different from $0$ one should probably not attach much intrinsic
significance to the details of (\ref{jocextra}), which are going to depend on the mentioned
issues of choice of Hopf-algebra basis and choice of ordering prescription,
but still (\ref{jocextra}) allows us to raise a significant point: a minimum-wavelenth
principle, when implemented in a quantum geometry with curvature, must take into
account the effects of redshift (illustrated in (\ref{jocextra}) through the
presence of the factor $e^{Ht_0}$).

Another feature which can be meaningfully discussed in relation to the broader
quantum-gravity/quantum-spacetime literature is the one of ``ultraviolet-infrared mixing".
It is expected that the short distance structure introduced for spacetime
quantization could (and perhaps should~\cite{cohenPRLuvir})
also affect the long-wavelength (infrared) regime~\cite{iruv1,iruv2}.
Some trace of infrared manifestations of the short-distance quantum-geometry structure
of our q-dS setup is present, although in rather implicit form,
in the result for the deformed comoving distance travelled by a massless particle
we obtained in Eq.~(\ref{eq:xqdS}) (and these infrared features will be exposed
more explicitly in reanalyses of (\ref{eq:xqdS}) which we discuss
later on in this manuscript).
We shall not dwell here on the quantitative details of these infrared
features, since they too should depend on
the mentioned
issues of choice of Hopf-algebra basis and choice of ordering prescription,
but we still feel that the possibility to investigate ultraviolet-infrared mixing
in a quantum geometry with curvature is very exciting.
In particular, flat-spacetime ultraviolet-infrared mixing, while attracting much interest
at the level of its technical description, is still confronted by severe
challenges of interpretation, because in a flat quantum spacetime the only
characteristic scale is the Planck scale, an ultraviolet scale that clearly
cannot on its own govern the onset of infrared features.
In our q-dS framework instead the curvature scalar and the Planck scale inevitably
cooperate, opening the way to more realistic descriptions of ultraviolet-infrared mixing.
The (energy) scale $H$ clearly could govern infrared features, and perhaps more excitingly
the framework also naturally allows for contemplating a role for the (squared-energy)
scale $H/L_p$, which is some sort of hybrid between the ultraviolet and the infrared
structure of the spacetime geometry, thereby potentially offering a particularly natural
candidate for the scale characteristic of the infrared side of the mechanism of
ultraviolet-infrared mixing.

\section{An application in astrophysics}\label{astrop}
The formalization developed in the previous section gives a definite picture
for the interplay of curvature and Planck-scale effects in dS-like (constant Hubble
parameter) quantum geometries, and clearly should also provide a meaningful first approximation
applicable to contexts in astrophysics that involve sources at relatively small distances
(small redshift, $z <1$), since then the analysis only involves rather small
time variations of the Hubble parameter.
And relevant for our thesis is the fact that
a much-studied possible implication of Planck-scale Hopf-algebra symmetries,
for which often results on $\kappa$-Minkowski theory provide at least part of
the motivation,
is the one of a dependence of the speed of massless particles on energy.
There is sizeable interest in particular in
studies~\cite{grbgac,astroBiller,kifune,ita,aus,gactp,tedOLDgood,emnPLB2009,gacSMOLINprd,fermiNATURE}
of this hypothesis
of energy-dependent speed for photons that exploit the
nearly ideal ``laboratory"
 provided by observations of gamma-ray bursts.
The sensitivity to ``in-vacuo dispersion" (fundamental energy dependence of
the speed of massless particles)
of gamam-ray-burst studies is rather significant, in spite of the
limitations
imposed by the fact that the source, the ``gamma-ray burster",
has intrinsic time variability~\cite{emnPLB2009,gacSMOLINprd,fermiNATURE}.
Previous related phenomenology work  assumed no interplay between
curvature and Planck-scale effects, so that from a Hopf-algebra perspective
it would amount to assuming that the energy dependence found in the $\kappa$-Minkowski
case should be added by hand to the analysis of particle propagation in
classical FRW spacetime.
We shall instead rely on the results reported
in the previous two sections to provide a preliminary
characterization of particle propagation in
a quantum (noncommutative) curved spacetime.

For these purposes we shall of course rely on
our analysis of $x_{q-dS}(t)$,
the comoving
distance travelled by a q-dS photon in a time interval $t$.
Since $L_p$ is small and, in the applications in astrophysics
that can be here of interest, $H$ is also small,
we can assume that our dependence of the parameter $w$ on $HL_p$ should be analyzed for
small values of $HL_p$. And for all the scenarios we considered for the relation $w=f(HL_p)$
one finds that $w$ is small when $HL_p$ is small. We can therefore rely
on an approximation of our result for $x_{q-dS}(t)$ valid\footnote{Notice that
Eq.~(\ref{approxIR}) formally has a singularity for $\Pi_P \rightarrow 0$.
This is an example of the infrared features mentioned in Subsection~\ref{jocperspe},
and, as we argued, one should not  necessarily interpret it
as an artifact of our approximations. It is known that
in noncommutative spacetimes (like our q-dS noncommutative spacetime)
such infrared issues may arise, and could be a meaningful manifestation
of the novel uncertainty principle for the localization of spacetime points
that a quantum spacetime predicts.
It is nonetheless reassuring that our results are however well behaved whenever $\Pi_P \ge H$
(and we are clearly not interested in particles with $\Pi_P < H$,
since $H_0 \sim 10^{-33}eV$).} for small $w$:
\begin{equation}
  x_{q-dS}^{(app)}(t)=\frac{e^{H t_0}-e^{-Ht}}{H}
  -w\,\frac{H^2 -4 e^{-2 H t}\Pi_P^2+4e^{2 H t_0}\Pi_P^2-e^{H( t + t_0)}H^2}{8 H^2 \Pi_P}\,.
  \label{approxIR}
\end{equation}

In the ideal case of two photons emitted simultaneously by a very compact source,
a photon with momentum $p$ and a photon with momentum $\ll p$,
we shall assume that the soft photon is detected at time $t=0$
while the photon of momentum $p$ is detected at some time $t=\delta_{qdS}$.
The two photons would have covered the same comoving distance, which, since for the soft
photon we can neglect quantum-spacetime effects~\cite{emnPLB2009,gacSMOLINprd},
we can denote by  $x_{dS}(0)$. This allows us to derive the delay time $\delta_{qdS}$
from the following equation:
\begin{equation}
 x_{dS}(0)=x_{q-dS}^{(app)}(\delta_{qdS})|_{\Pi_p=-pe^{2 H \delta_{qdS}}}~.
\end{equation}
Working in leading order in $\delta_{qdS}$ one then obtains
\begin{equation}
 \delta_{qdS}=\frac{\left(-H^2 +4 p^2-4 e^{2 H t_0} p^2+e^{H t_0} H^2 \right) w}{8 H^2 p }
 \simeq \frac{p(1-e^{2 H t_0}) w}{2 H^2} ~,
 \label{era79}
\end{equation}
where on the right-hand side we took into account
of the smallness of the values of $H$ that are relevant for gamma-ray-burst studies
($H\sim 10^{-33}  \text{eV}$).

The relevance for our thesis of this result (\ref{era79}) originates from the
explicit dependence of $\delta_{qdS}$ on $H$ and perhaps even more significantly
from the implicit dependence on $H$ contained in $w$. The Planck length $L_p$
only appears in $\delta_{qdS}$ through $w$, and therefore the interplay between curvature
and Planck-scale effects is very significant. The part of this interplay codified in the fact
that necessarily $w$ must be written in terms of $H L_p$ is evidently a robust feature of
our framework, qualitatively independent of the mentioned issues concerning the choice
of Hopf-algebra basis and the choice of ordering prescription. The residual dependence on $H$
(such as the factor $e^{2 H t_0}$) is instead a more fragile aspect of our analysis,
but still indicative of the type of qualitative features
that one in general should expect.

Also notice that (\ref{era79}) reflects the requirement $\alpha \geq 1$
that emerged in the analysis we reported in the previous section, for cases
with $w \simeq (HL_p)^\alpha$.
In fact, for $\alpha < 1$ the $H \rightarrow 0$
limit of (\ref{era79}) is pathological (whereas no pathology arises for $\alpha \geq 1$).

\section{dS slicing of a FRW Universe }\label{classlice}
Our results on a quantum-dS spacetime dual to a quantum/Hopf algebra of symmetries
are of potential relevance also for the astrophysics of distant sources ($z \gtrsim 1$)
and for cosmology, but only
in a rather indirect way. We shall argue that taking as starting point
our dS-like quantum spacetime represents an advantage of perspective
with respect to the most common strategy adopted so far in the study of
Planck-scale effects in astrophysics and cosmology, which relies on taking as starting
point results on the quantization of Minkowski-like spacetimes. The most evident
advantage is of course due to the fact that at least we can make use of
some intuition on the interplay between curvature and Planck-scale effects,
which was indeed the primary motivation for our study.

Before actually applying our strategy to quantum-spacetime contexts, we find useful
to devote this Section to a sort of test of this strategy in a classical-spacetime
context, where of course it is easier for us (and for our readers) to confindently
assess its efficacy.
Specifically, in this Section
we introduce, at the classical level,
an approximation of a FRW solution in which the time axis is divided
in small $\Delta t$ intervals and in each $\Delta t$ interval the FRW Universe
is approximated by a corresponding ``slice" of the
dS space-time, and we compare our findings to the corresponding results
one obtains by using analogously ``Minkowski slices" (which is the classical-spacetime version
of the strategy used in most studies of
Planck-scale effects in astrophysics and cosmology).
It is obvious {\it a priori} that both ``slicing strategies" must converge to a faithful
description of the FRW results in the limit of very detailed slicing (locally, in an
infinitesimal neighborhood of a point in FRW both approximations are exact of course).
But it is interesting to find confirmation that even at the classical level
the dS slicing, through its ability of codifying information on curvature,
is a better tool of approximation.

It is sufficient for our purposes to contemplate the case
of the FRW solution whose line-element is
\bea
&& ds^2=dt^2 - a^2(t)\, {d\vec{x}}^2 ~.\label{dsrad}
\eea
In order to provide explicit formulas it is useful for us to fix the
conformal factor $a(t)$, and we take as illustrative example the
case $a(t)=a_{rad-3D}(t) \equiv (t/t_0)^{2/3}$,
where $t_0$ is a normalization time\footnote{In this and the next
section we set t = 0 at the Big
Bang time.}.
While in this section this choice of $a(t)$ is as good as any other possible illustrative
example, but we shall again adopt
this choice of $a(t)$ in the next section, which is where we
report the aspects of our analysis that are potentially relevant for cosmology.
The choice $a(t)= (t/t_0)^{2/3}$ is adopted there by the desire to contemplate
the first radiation-dominated instants of evolution of the Universe,
while taking into account that most of our insight on the theory side,
and particularly the {\it ansatz} reported in Eq.~(\ref{wlqg}),
originated in 2+1D theories: indeed
 one finds
that $a(t)= (t/t_0)^{2/3}$
in a radiation-dominated 2+1D FRW Universe.

Let us then split the time interval $ \{ 0,t_F \} $,
in $n$ small intervals $\Delta t = t_F /n$, with $n\in \mathbb{N}$. In each $k$-th
interval $\Delta t$  we consider a corresponding ``slice"
of dS space-time, with time variable $t'\in\left[0,\Delta t\right]$, spatial coordinate $\vec {x}'$ and
with line element given by
\be
ds_k^2= dt'^2 -A^2_k e^{2H_k t'} {d \vec{x}'}^2\,.
\label{joclabel1}
\ee
Here $H_k$ is related to the square root of the ``effective cosmological constant''
relative to the $k$-th interval, and the additional
parameter, the scaling constant $A_k$, must also appropriately match some FRW requirements.
Specifically, in order to
match the FRW evolution given by $a(t)$,
we must fix $A_k$ and $H_k$, in each $k$-th interval of duration $\Delta t$,
through the following requirements:
\begin{equation}
 a(t=m \Delta t+t')=\sum_{k=1}^{n} A_k \, e^{H_k t' } \,
\theta\!\qs (k+1) dt\!-\! t \qd \theta\!\qs t\!
-\! k \Delta t \qd, \label{atdiscr}\\
\end{equation}
where the product of the two heavyside functions selects the $m$-th term in the summation.

For $a(t)=a_{rad-3D}(t) \equiv (t/t_0)^{2/3}$ this leads to:
\bea
A_k&=& \left(\frac{k \Delta t}{t_0}\right)^{2/3}\,,\\
H_k
&=& \f{2}{3\Delta t} \ln \ts  \f{k+1}{k} \td,\label{Hi}
\eea
where we have imposed $A_k=a_{rad-3D}(k\Delta t)$
and $a_{rad-3D}(k\Delta t+t')|_{t'=\Delta t}=a_{rad-3D}((k+1)\Delta t)$.

With the approximated $a(t)$ one can calculate the physical distance
travelled by a FRW photon in a time $t_F=n_F\, \Delta t$, which is given by\footnote{Notice
that the lowest value of $k$ is $k = 1$ (rather than $k=0$ as one could
perhaps naively imagine). This, as shown later in this section,
ensures that our ``slicing procedure" converges (in the vanishingly-thin-slice limit)
to a faithful descritpion of the physical distance
travelled by a FRW photon. If one naively added the term with $k=0$
the result would be pathologically divergent,
since
$\lim_{k\to 0} \,{(k \Delta t)^{-{2}/{3}}}\,\left[{3 \Delta t}/{  \ln\ts 1 + \f 1k \td^2  } \right]\ts  1-{1}/{\sqrt{1+\f 1k}} \,\td = \infty~.$
}
\bea
l_{FRW}^{(dS)}(t_F)& =&  a(t_F) \int_0^{t_F} \f{dt}{a(t)}=a(t_F) \sum_{k=1}^{n_F}\int_0^{\Delta t} \f{dt'}{a(k\Delta t+t')}=\nn\\
&=&  (t_F)^{\frac{2}{3}}\, \sum_{k=1}^{n_F} \f{1}{(k \Delta t)^{\frac{2}{3}}}
 \f{1-e^{-H_k \Delta t}}{H_k}=(t_F)^{\frac{2}{3}} \, \sum_{k=1}^{n_F }\f{1}{(k
\Delta t)^{\frac{2}{3}}} \qs  x_{dS}(\Delta t) \qd_k
 ~,
\label{qdSpath}
\eea
where for $\qs  x_{dS}(\Delta t) \qd_k$ we use Eq.~(\ref{eq:xdS})
(with $t_0=0$ and $H=H_k$  given by Eq.~(\ref{Hi})).

Within our approximation, Eq.~(\ref{qdSpath}),
the distance travelled by the FRW photon  is expressed in terms of the formula
that gives the comoving
distance travelled by a dS photon.
In Planck units, if $t_F=1$ the exact distance travelled by the FRW photon
is $ l_{FRW}(1)=3 $, which also encodes the ``horizon
paradox" (the distance travelled
by the FRW photon starting from $t=0$ up to a time $t_F$
is always smaller than what would be needed in order to achieve agreement
with the observed isotropy of the CMBR).
We of course repoduce faithfully this feature in our ``dS slicing of FRW"
in the limit of vanishingly small slices:
\be
\lim_{\Delta t \to 0} l_{FRW}^{(dS)}(1)=3\,.
\ee

In order to test the accuracy of our approximation (\ref{qdSpath})
in the evaluation of the distance travelled by a FRW photon,
it is useful to compare $l_{FRW}^{(dS)}(t_F)/a(t_F)$ and $l_{FRW}/a(t_F)$.
For this purpose we observe that
\begin{eqnarray}
\Delta l^{(dS)} & \equiv & \frac{l_{FRW}^{(dS)}(t_F)}{a(t_F)}-\frac{l_{FRW}}{a(t_F)}
  =t_0^{\frac{2}{3}}\;\;\sum_{k=1}^{n_F} \f{1}{(k \Delta t)^{\frac{2}{3}}}
   \f{1-e^{-H_k \Delta t}}{H_k}-t_0^{\frac{2}{3}}\sum_{k=1}^{n_F}\int_{\Delta
    t k}^{\Delta t (k+1)}\frac{dt'}{{t'}^{\frac{2}{3}}}\nonumber \\
&=& 3\,  t_0^{\frac{2}{3}}\;\;\sum_{k=1}^{n_F}{\Delta t}^{\frac{1}{3}} \left[
 \frac{1}{2\log{(1+\frac{1}{k})}}\left(\frac{(\frac{1}{k}+1)^{\frac{2}{3}}
 -1}{(k+1)^{\frac{2}{3}}}\right)-(k+1)^{\frac{1}{3}}+k^{\frac{1}{3}}   \right]
~, \label{eq:FRWvsDS}
\end{eqnarray}
from which one easily infers the exact agreement achieved
for $\Delta t\rightarrow 0$.

It is interesting for us
to examine how our approximation, for finite $\Delta t$, compares
with an analogous approximation based on ``Minkowski slices".
In a Minkowski slicing the line element in each $k$-th interval has the form:
\begin{equation}
 ds_k^2=dt'^2-B_k^2 dx'^2,
\end{equation}
with $B_k=({k\Delta t}/{t_0})^{2/3}$. And the associated
natural approximation
of $l_{FRW}(t_F)$ is
\begin{equation}
 l_{FRW}^{(M)}(t_F)=t_F^{\frac{2}{3}}\sum_{k=1}^{n_F}\frac{\Delta t^{\frac{1}{3}}}{k^{\frac{2}{3}}}.
\end{equation}
The accuracy of this Minkowski-slicing approximation can be inferred from examining
\begin{equation}
\Delta l^{(M)} \equiv \frac{l_{FRW}^{(M)}(t_F)}{a(t_F)}-\frac{l_{FRW}}{a(t_F)}
=t_0^{\frac{2}{3}}\sum_{k=1}^{n_F}\Delta t^{\frac{1}{3}}\left[\frac{1}{k^{\frac{2}{3}}}-3(k+1)^{\frac{1}{3}}
+3k^{\frac{1}{3}}\right]\label{eq:FRWvsM}
\end{equation}
And it is noteworthy that
each $k$-th element of this summation is bigger than the corresponding $k$-th element of
the dS-slicing case (\ref{eq:FRWvsDS}).
Since all the terms in these summations are positive,
one then concludes that
the total difference between the physical distance
calculated exactly in FRW and the one calculated though
Minkowski slicing is always bigger than corresponding difference
between the exact distance and the one calculated through dS slicing.

\section{q-dS slicing of a q-FRW Universe} \label{sec:qFRW}
In this section we contemplate a possible application
of our scheme of analysis in a regime where curvature
does take large values and therefore the interplay between Planck-scale and curvature
can be particularly significant. This is the context of studies of the propagation
of massless particles in the early Universe, focused primarily on its implications
for causality.
We shall proceed following a strategy which is inspired by the
observations we reported in the previous section.
Since the Hopf-algebra/noncommutative-spacetime literature does not offer candidates\footnote{Within
the Loop-Quantum-Gravity approach there have been recent proposals~\cite{lqc1,lqc2,lqc3}
of quantum-geometry descriptions of the early Universe. The type
 of issues that we are here concerned with has not yet been studied within this
 Loop-Quantum-Gravity approach, but insightful results have been obtained for example
 in investigations of the possibility
that such quantum geometries may be suitable
for a description of the early Universe that is free from a $t=0$ singularity.}
for a ``q-FRW Universe" (a FRW-like quantum spacetime)
we shall rely on the assumption that propagation of massless particles in
such a spacetime admits approximation in terms of ``q-dS slicing",
so that we can once again rely on our result for
 the q-dS comoving distance.

Clearly for the application to the early Universe that we propose in this section
it is more difficult to gauge the size of the inaccuracies and fragilities
introduced by our approximations and choices of ordering prescriptions.
We still expect our analysis to display the qualitatively correct nature of the
interplay between curvature and Planck scale in the early Universe (according
to the general framework ispired by $\kappa$-Minkowski and q-dS),
but quantitatively the approximations we produce may well eventually
turn out to be rather poor.
Still we feel that the lessons learned through our analysis
are valuable, especially in light of the
fact that the first pioneeristic $\kappa$-Minkowski-inspired studies
of the early Universe produced so far (see, {\it e.g.} Refs.~\cite{infla1,infla2,infla3,maguREV}
and references therein) completely neglect the possibility of interplay
between curvature and Planck-scale effects.

Our case for the significance of the
interplay between curvature and Planck-scale effects in the early Universe
is based on a description of
 the motion of a photon in a
quantum FRW Universe obtained as a summation
of terms given by our proposal for the q-dS comoving distance.
And our ``q-dS slicing" assumes a description of $a(t)$ which,
as already done and motivated in the previous section, is the one appropriate
for a radiation-dominated era in 2+1D cosmology, $a(t)=a_{rad}(t)= (t/t_0)^{2/3}$.
As long as $t$ is small but still $t>1$ in Planck units, $a_{rad}(t)$
is a good approximation
of the evolution of such a 2+1D Universe. For $t \lesssim 1$
one clearly expects new physics
to come into the picture. But for our exploratory purposes we
choose to simply adopt $a_{rad}(t)$
even for values of $t$ all the way down to  $t=0$,
where according to the classical setting
the primordial Big-Bang point, a singularity in the Riemann tensor, is found.
This setup will allow us in particular to produce
observations that are directly relevant for a research
programme~\cite{infla1,infla2,infla3,maguREV}
which explores the possibility of introducing Planck-scale-modified laws of propagation
as a way to replace inflation in solving some of the cosmological paradoxes.
We shall not ourselves dwell on whether or not these proposals are promising, but we rather
intend to provide
new tools that could play a role in future investigations of the effectiveness of these proposals.

In light of these preliminary
considerations our starting point clearly must be equation (\ref{qdSpath}),
rewritten for the case of q-dS slicing (rather than the original dS slicing)
through the substitution
\be
\qs x_{dS}(\Delta t) \qd_k \longmapsto \qs x_{q-dS}(\Delta t) \qd_k\,.
\ee
For  $\qs x_{q-dS}(\Delta t) \qd_k$ we can rely on Eq.~(\ref{eq:xqdSpe}),
with $H=H_k$ and $t_0=0$:
\begin{equation}
 [x_{q-dS}(t)]_k=\frac{\sqrt{2 w \tanh{(\frac{w}{2})}}}{w H_k
(1-e^{\frac{-w\Pi_E}{H_k}})} \ln{[Z_k] }\,,
\end{equation}
where $Z_k$ can be written as
$$Z_k=  \cosh{ (\frac{w}{2}) }  e^{\frac{w \Pi_E}{H_k}}
-\sqrt{  (1-\cosh{(\frac{w}{2})e^{\frac{w \Pi_E}{H_k}})^2
+\cosh{(\frac{w}{2})}( 1-e^{ \frac{-w\Pi_E}{H_k} }  )^2  (e^{-2 H_k t}-1) }  } \,.$$

Of course, one also needs a procedure for ``slice matching":
our $ \qs x_{q-dS}(dt) \qd_k $
depends on $\Pi_E$, which is a conserved quantity in our  q-dS framework,
but clearly would not be a good conserved charge in a quantum FRW Universe.
This issue of the determination of $\Pi_E$ in each $k$-th
interval is clearly related to the scaling of energy
in a quantum FRW Universe. This is a key point where, because of the unavailability of
a formulation of quantum FRW spacetime, we can only proceed by adopting
a plausible {\it ansatz}, thereby loosing control on the quantitative accuracy of our estimates.
However, as stressed already in other points of this manuscript, our primary objectives
are not of detailed quantitative nature. Specifically, in this section on implications
for the early Universe we simply want to provide some support for our main point
that the interplay between curvature and Planck-scale effects can be particularly significant
at high curvature. And we base our thesis not on the details of what we find for one
particular choice of $w(HL_p)$ (the dependence of the deformation parameter $w$ on $H L_p$),
 but rather on the comparison between the results obtained
for two different but related choices of $w(HL_p)$. In light of this our
(qualitative) findings
are relatively insensitive to changes of the choice of {\it ansatz} used
to fix $\Pi_E$ in each $k$-th interval.

The {\it ansatz} we adopt is inspired by the
scaling of the energy $E_{in}=E_{F} [a(t_{F})/a(t_{in})]$ in a classical FRW universe.
We further estimate the scale factor by considering the FRW radiation
dominated era $a_{rad}(t)=(t/t_0)^{2/3}$.
This in turn leads us to assuming that at the initial time of each k-th
interval, within our ``q-dS slicing", $\left[\Pi_E\right]_k$ should
be tentatively described as follows:
\be
\qs \Pi_E \qd _k \varpropto \f{E_F t_F^{2/3}}{(k\,\Delta t)^{\frac{2}{3}}} ~,\label{FRW2scales}
\ee
which has already been specialized to our illustrative example of time dependence
of the scale factor $a(t)= (t/t_0)^{2/3}$.

With this {\it ansatz} we specified the only unknown of our ``slicing procedure"
for the description of a quantum FRW spacetime in terms of our findings for
the q-dS spacetime. In principle this can be used for preliminary calculations
of physical distances over which a photon propagates in a certain chosen time interval.
But it appears that some striking indications of the curvature dependence of Planck-scale effects,
our main objective here, can be found even without embarking in a numerical analysis.
We can do this by a comparison of two of the cases for the relationship between $w$, $H$ and $L_p$
which we motivated in Section~\ref{q-dS}, specifically the case\footnote{Note that in
Section~\ref{q-dS}, because of the objectives of that part of our analysis,
we were satisfied to consider generically the possibility $w \propto H L_p$,
while here our desire to establish a more precise connection
with the alternative choice $w=\frac{2\pi}{2+1/H L_p}$
leads us to contemplate specifically the case $w= 2 \pi H L_p$.
In this way we arrange a comparison that is particularly insightful
since $\frac{2\pi}{2+1/H L_p} \approx 2 \pi H L_p$ for small $H$,
while $\frac{2\pi}{2+1/H L_p} \ll 2 \pi H L_p$ for $H \gg 1/L_p$.} $w= 2 \pi H L_p$
and the case $w=\frac{2\pi}{2+1/H L_p}$.
This is interesting because for small values of $H$ one has $\frac{2\pi}{2+1/H L_p} \simeq 2\pi H L_p$
but for large values of $H$ one has that $\frac{2\pi}{2+1/H L_p} \neq 2\pi H L_p$,
and therefore the formula $w=\frac{2\pi}{2+1/H L_p}$ (which, as mentioned in Section~\ref{q-dS},
is one of the few all-order formulas of this type that finds some support at least in
one quantum-gravity approach) is itself a probe of the possible relevance of the curvature dependence
of Plancks-scale quantum-spacetime effects.

To expose the sought curvature dependence of the Planck-scale effects
it suffices to examine
the $\Delta t\rightarrow 0$ limit of the generic $k$-th term in our descripion for the physical
distance in q-FRW, which is given by
\be
\lim_{\Delta t\to 0}\frac{ t_F^{\frac{2}{3}}\,[x_{q\!-\!dS}(\Delta t)]_k}{(k \Delta t)^{\frac{2}{3}}}\,.
\ee

For the case $w=2 \pi H L_p$,
also taking into account the behaviour
of $[\Pi_E]_k$ given in (\ref{FRW2scales}), one finds that for small $\Delta t$
(``fine slicing')
 \be
 \frac{ t_F^{\frac{2}{3}}\,[x_{q\!-\!dS}(\Delta t)]_k}{(k \Delta t)^{\frac{2}{3}}}
 \simeq \frac{ t_F^{\frac{4}{3}}}{ k^{\frac{4}{3}}} 2\sqrt{\pi L_p} E_F \frac{(\Delta t)^{1/6}}{(2/3 \log(1+\frac{1}{k}))^{3/2}}\,.
 \label{caso1}
\ee
Looking at the same quantity for the case $w=\frac{2\pi}{2+1/H L_p}$
one finds instead, again for small $\Delta t$,
 \be
 \frac{ t_F^{\frac{2}{3}}\,[x_{q\!-\!dS}(\Delta t)]_k}{(k \Delta t)^{\frac{2}{3}}}
 \simeq  \frac{ t_F^{\frac{2}{3}}}{k^{\frac{2}{3}}} \frac{\sqrt{\tanh(\pi/2)}\left(\cosh(\pi/2)\right)^2}{\cosh(\pi/2)-1}\frac{3}{ \sqrt{2\pi} \log(1+\frac{1}{k})} (\Delta t)^{1/3}
 \label{caso2}
\ee

The fact that curvature affects the quantum properties of
our q-dS spacetime, in ways that mainly originate from the dependence of $w$ on $H$,
is here reflected in the fact that two different but related {\it ans\"atze}, $w=2 \pi H L_p$
 and $w=\frac{2\pi}{2+1/H L_p}$
produce different descriptions
of the physical distances over which a photon propagates in a given time interval.
The main differences are encoded in the different dependence on the ``slice label" $k$,
since $k$ labels primarily $H_k$ and the two scenarios we are comparing
have different dependence of $w$ on $H$ ({\it i.e.} $w_k$ on $H_k$).
Also notice that associated to this different $k$ dependence one also finds
a different dependence on $\Delta t$, the ``thickness" of the slice in the time direction.
This is ultimately where the most profound causality-relevant implications
should be found, since then by taking the $\Delta t \rightarrow 0$ limit of sums
of terms of this sort one would compute the overall distance travelled
by the photon. For some choices of the dependence of $w$ on $H$ and $L_p$
one should expect to even find that
the physical distance
over which a photon propagates in a given finite interval can diverge,
as essentially assumed (without however considering a possible role for curvature)
in Refs.~\cite{infla1,infla2,infla3,maguREV}.
We shall not dwell on these possibilities here since we perceive
in the limited scopes
of our analysis (particularly for what concerns choice of ordering prescription
and the {\it ansatz} (\ref{FRW2scales}))
an invitation to be prudent at the level of quantitative predictions,
but we do feel that our findings are robust
for what concerns the significance of the interplay between curvature
and Planck-scale effects in early-Universe cosmology, here qualitatively
exposed by comparing Eqs.~(\ref{caso1}) and (\ref{caso2}).

\section{Outlook}\label{out}
The interplay between curvature and Planck-scale effects, whose significance was
here robustly established, could well gradually acquire a key role in quantum-gravity-inspired
studies in astrophysics and cosmology.
Within a fully specified model of course one would have a definite
prediction for this interplay, which could be very valuable
from a  phenomenology perspective, since it would provide a more distinctive
characterization  of the physical implications of the model.
This could be exploited for example in cases where
the candidate quantum-gravity effect of interest is subject to ``competition"
by other new-physics proposals (see, {\it e.g.}, Ref.~\cite{dmINVACUODISP}).
While these could indeed be valuable opportunities produced by the interplay
between curvature and Planck scale, it is likely that at least for
some quantum-gravity/quantum-spacetime models (depending on the specific form
of interplay that a given model will predict) this interplay will also introduce
some challenges. For example, the fate of causality in quantum spacetime
is of course a major concern, and the preliminary analysis we reported
in Section~\ref{sec:qFRW} suggests that the implications for causality
of some of these quantum-gravity-inspired models
might be amplified in contexts involving large curvature,
such as the description of the
early Universe.

The model dependence of these challenges and opportunities
should be explored within the q-dS framework and other possible formalizations
of scenarios with curved quantum spacetimes. For the q-dS framework we here exposed
some key aspects of this model dependence, which mainly concern possible
ambiguities originating from changes of
 ordering prescription and nonlinear redefinitions of the basis of
Hopf-algebra generators. Of course, it would be desireable to show that the physical predictions
of the q-dS framework do not depend on these apparently arbitrary choices, and some
results obtained for flat quantum spacetimes (and their Hopf algebras of symmetries)
provide encouragement~\cite{kappa1, theta} for this hope. But also
in this respect the presence of curvature may introduce some challenges, which should be addressed
in dedicated studies.

\section*{Acknowledgements}
We gratefully acknowledge insightful conversations with M.~Matassa and G.~Rosati.
The work of G. Amelino-Camelia and A. Marcian\`o was supported in part
by grant RFP2-08-02 from The Foundational Questions Institute (fqxi.org).
A. Marcian\`o also gratefully acknowledges ``Universit\`a di Roma La Sapienza''
and ``Fondazione Angelo Della Riccia'' for financial support during the period
in which this work has been carried out.
The work of G. G. and A. M. was supported by ASI contract I/016/07/0 "COFIS".

{}

\end{document}